\documentclass[aps, pre, twocolumn,preprintnumbers,amsmath,amssymb,longbibliography,showkeys]{revtex4-2} %longbibliography

\usepackage{graphicx}% Include figure files
\usepackage{dcolumn}% Align table columns on decimal point
\usepackage{bm}% bold math
\usepackage{xcolor}
\usepackage{amssymb}

\usepackage{caption, subcaption}
\usepackage{float}

\usepackage[normalem]{ulem}

\usepackage{setspace}
\usepackage[normalem]{ulem}
\usepackage{xcolor}
\usepackage{bm}
\usepackage{hyperref}
\begin{document}
	
	%\preprint{APS/123-QED}
	
	%\title{
		%Nonequilibrium thermodynamics of a particle driven by time-delayed feedback
		%}
	\title{
		Heat production in a stochastic system with nonlinear time-delayed feedback
	}
	\author{Robin A. Kopp}
	\email{r.kopp@tu-berlin.de}
	\author{Sabine H. L. Klapp}
	\email{sabine.klapp@tu-berlin.de}
	\affiliation{%
		Institut für Theoretische Physik, Technische Universität Berlin, Hardenbergstr. 36, D-10623 Berlin, Germany\\
	}
	
	\date{\today}% It is always \today, today,
	
	\begin{abstract}
			Using the framework of stochastic thermodynamics we study heat production related to the stochastic motion of a particle driven by repulsive, nonlinear, time-delayed feedback.
			Recently it has been shown that this type of feedback can lead to persistent motion above a threshold in parameter space [Physical Review E \textbf{107}, 024611 (2023)]. Here we 
			investigate, numerically and by analytical methods, the rate of heat production in the different regimes around the threshold to persistent motion. We find a nonzero average heat production rate, $\langle \dot{q}\rangle$, already below the threshold, indicating the nonequilibrium character of the system even at small feedback. In this regime, we compare to analytical results for a corresponding linearized delayed system and a small-delay approximation which provides a reasonable description of $\langle \dot{q}\rangle$ at small repulsion (or delay time).
			Beyond the threshold, the rate of heat production is much larger and shows a maximum as function of the delay time.
			In this regime, $\langle \dot{q}\rangle$ can be approximated by that of a system subject to a constant force stemming from the long-time velocity in the deterministic limit. The distribution of dissipated heat, however, is non-Gaussian, contrary to the constant-force case.
	\end{abstract}
	
	%Use showkeys class option if keyword
	\maketitle
	%\vspace{-15pt}
	%\tableofcontents
	%\newpage
	%%%%%%%%%%%%%%%%%%%%%%%%%%%%%%%%%%%%%%%%%%%%%%%%%%%%%%%%%%%%%%%%%%%%%%%%%%%%%%%%%%%%%%%%%%%%%%%%%%%%%%%%%%%%%%%%%%%%%%%%%%%%%%%%%%%%%%%%%%%%%%%
	%
	%
	%
	%
	%%%%%%%%%%%%%%%%%%%%%%%%%%%%%%%%%%%%%%%%%%%%%%%%%%%%%%%%%%%%%%%%%%%%%%%%%%%%%%%%%%%%%%%%%%%%%%%%%%%%%%%%%%%%%%%%%%%%%%%%%%%%%%%%%%%%%%%%%%%%%%%
	\section{Introduction}
	Feedback control \cite{bechhoeferControlTheoryPhysicists2021,bechhoefer_feedback_2005} is a concept used in many fields of sciences, ranging from chaos control \cite{scholl_eckehard_handbook_2007,pyragas_continuous_1992,pyragas_control_1995,tsui_control_2000} and engineering \cite{bechhoeferControlTheoryPhysicists2021}, where feedback has its origin,
	to applications in laser systems \cite{bielawski_controlling_1994,marino_pseudo-spatial_2017}, quantum systems \cite{carmele_single_2013}, trapping
	 \cite{junHighPrecisionTestLandauer2014a,kumarNanoscaleVirtualPotentials2018}
	%lopezRealizationFeedbackControlled2008
	and feedback cooling \cite{gieselerSubkelvinParametricFeedback2012,gieseler_non-equilibrium_2015} of nanoparticles. While feedback is often assumed to be applied instantaneously, this assumption does not hold for many real-world scenarios. 
	Indeed, in many experiments, feedback mechanisms require a finite time for collecting and processing information (e.g., through a camera), and subsequently manipulating the dynamics (see, e.g., \cite{franzl_active_2020}). 
	More generally, delay is often not negligible in chemical reactions \cite{thanh_efficient_2017} and active matter \cite{holubec_finite-size_2021,pakpourDelayinducedPhaseTransitions2024},
	as well as in macroscopic contexts such as traffic flow \cite{kerner_introduction_2009,yuDensityWavesTraffic2010,sipahiStabilityTrafficFlow2008}, robotic systems \cite{mijalkov_engineering_2016}, and financial markets \cite{stoica_stochastic_2004,callen_accounting_2013}.
	Importantly, time delay can also have a {\em constructive} effect. In particular, in the framework of feedback control, it can be used to stabilize unstable orbits (Pyragas control) \cite{pyragas_continuous_1992,pyragas_control_1995} and to induce 
	oscillations \cite{erneux_applied_2009-1} or other new types of motion \cite{khadkaActiveParticlesBound2018}.

	Recently, there is increasing interest in applications of time-delayed feedback to colloidal and active systems that are, intrinsically, stochastic due to thermal fluctuations. Early applications have focussed, e.g., 
	on the manipulation of transport properties \cite{lopezRealizationFeedbackControlled2008,loos_delay-induced_2014}. Since then, there have been substantial experimental advances \cite{franzl_active_2020,loffler_behavior-dependent_2021,fernandez-rodriguez_feedback-controlled_2020-1,bell-davies_dynamics_2023} concerning real-time tracking 
	(e.g., using deep learning algorithms) and control involving time delay (e.g., by photon nudging \cite{muinos-landinReinforcementLearningArtificial2021,wangSpontaneousVortexFormation2023}, optical tweezers
	%wallin_stiffer_2008
	\cite{wallin_stiffer_2008,gavrilovFeedbackTrapsVirtual2017}, adaptive light fields \cite{fernandez-rodriguez_feedback-controlled_2020-1,bell-davies_dynamics_2023}, etc.) of individual or even multiple colloidal particles~\cite{franzl_active_2020,chenPersistentResponsiveCollective2024}. 
	There are also recent experiments involving time-delayed feedback in the underdamped regime \cite{debiossac_thermodynamics_2020}.
	In parallel, theoretical studies have addressed both, the dynamical behavior of single particles with different types of time-delayed feedback \cite{craigEffectTimeDelay2008,gernertFeedbackControlColloidal2016,khademDelayedFeedbackControl2019a}, as well as collective behavior \cite{tarama_traveling_2019,kopp_spontaneous_2023,holubec_finite-size_2021,pakpourDelayinducedPhaseTransitions2024}.

	In the present study, we consider an overdamped stochastic (colloidal) system subject to a repulsive, nonlinear feedback force with time delay. It has been shown that this type of feedback, when applied to a single particle, can induce persistent motion \cite{kopp_persistent_2023-1,bell-davies_dynamics_2023}. Furthermore, in many-particle systems, repulsive feedback can lead to rich collective behavior~\cite{tarama_traveling_2019,kopp_spontaneous_2023}, such as Vicsek-like~\cite{vicsek_novel_1995} velocity ordering~\cite{kopp_spontaneous_2023}. 
	Importantly, repulsive time-delayed feedback can also be realized experimentally~\cite{bell-davies_dynamics_2023}.

	Here we are concerned, in particular, with thermodynamic signatures of the feedback-induced onset of persistent motion of a single particle. To this end we use the framework of stochastic thermodynamics~\cite{sekimoto_langevin_1998,seifert_stochastic_2012-1,seifertStochasticThermodynamicsPrinciples2018a,peliti_stochastic_2021}. In doing this, we are facing two major challenges.
	First, a finite time delay renders the particle's dynamics non-Markovian: due to the history dependency the problem becomes, formally, infinite-dimensional (for earlier studies of this type of stochastic motion, see \cite{guillouzic_small_1999-1,giuggioli_fokkerplanck_2016-1}). Due to this complication, "standard" concepts of stochastic thermodynamics 
	have to be reconsidered. This concerns, in particular, the total entropy production rate which has been thoroughly studied for underdamped systems with time delay
	\cite{munakataEntropyProductionFluctuation2014,rosinberg_stochastic_2015,rosinberg_stochastic_2017,debiossac_thermodynamics_2020,rosinbergFluctuationsDynamicalObservables2024}. For linear feedback forces, some of these problems can be circumvented by introducing auxiliary variables (Markovian embeddeding) \cite{loosIrreversibilityHeatInformation2020,loosMediumEntropyReduction2021,antary_matrix_2024}. This brings us to the second main challenge related to our present system, that is, nonlinearity. Indeed, most explicit analytical results for the thermodynamics of systems with time delay have been obtained for linear systems only \cite{munakata_linear_2009,munakataEntropyProductionFluctuation2014,rosinberg_stochastic_2015,rosinberg_stochastic_2017,rosinbergFluctuationsDynamicalObservables2024} (see, however, \cite{loos_heat_2019}). Given these difficulties, a full analytical study of the nonequilibrium stochastic thermodynamics of nonlinear time-delayed feedback seems currently out of reach. 

	Here we restrict ourselves to essentially one quantity characterizing the nonequilibrium thermodynamics, that is, the average rate of heat production, as well as the distribution of dissipated heat  (which is well studied in various Markovian model systems, see, e.g., \cite{speck_distribution_2004,speck_distribution_2007-1,imparato_work_2007}). Our main question is how these properties change upon crossing the parameter threshold related to the onset of persistent motion.
	Indeed, the question of how thermodynamic notions reflect (phase) transitions and bifurcations, even in single-particle systems, is a topic of current interest \cite{herpichCollectivePowerMinimal2018,borthneTimereversalSymmetryViolations2020a,grandpreEntropyProductionFluctuations2021,yuEnergyCostFlocking2022,ferrettiSignaturesIrreversibilityMicroscopic2022a,fernandezNonequilibriumDynamicsEntropy2024}. 
	Another relevant question is to which extent heat production in the nonlinear system differs from that in the linear case \cite{munakata_linear_2009,loos_heat_2019,loosIrreversibilityHeatInformation2020}. To investigate these questions we perform numerical 
	simulations as well as analytical calculations for several limiting cases. These include small feedback strength, small delay, and strong forcing leading to a constant velocity in the deterministic case.

	The remainder of this paper is structured as follows:
	In Sec.~\ref{sec:model} we introduce the model and describe the general behavior of the system. Then, splitting the parameter space into a regime below and above the persistence threshold, we present in Sec.~\ref{sec:analytical_considerations} various analytical considerations concerning the heat production rate. Here we present, first, a discussion of the linearized problem corresponding to small feedback (Sec.~\ref{sec:linear_case}), including the question of the existence of a stationary state (Sec.~\ref{sec:stable_system}), and the limit of small delay (Sec.~\ref{sec:small_delay_approx}). 
	Second, we consider heat production in the constant-velocity approximation (Sec.~\ref{sec:persistent_motion}). Numerical results and comparisons with the analytical results are presented in Sec.~\ref{sec:numerical_results}. Here we discuss dependencies of the average heat production rate on parameters, as well as properties of the distribution of dissipated heat.
	%%%%%%%%%%%%%%%%%%%%%%%%%%%%%%%%%%%%%%%%%%%%%%%%%%%%%%%%%%%%%%%%%%%%%%%%%%%%%%%%%%%%%%%%%%%%%%%%%%%%%%%%%%%%%%%%%%%%%%%%%%%%%%%%%%%%%%%%%%%%%%%
	%
	%
	%
	%
	%%%%%%%%%%%%%%%%%%%%%%%%%%%%%%%%%%%%%%%%%%%%%%%%%%%%%%%%%%%%%%%%%%%%%%%%%%%%%%%%%%%%%%%%%%%%%%%%%%%%%%%%%%%%%%%%%%%%%%%%%%%%%%%%%%%%%%%%%%%%%%%
	%
	%Main text
	%
	\section{Model and thermodynamic notions\label{sec:model}}
	\subsection{Equations of motion and single-particle dynamics\label{sec:twoA}}
	We consider a non-Markovian stochastic system composed of a colloidal particle moving under the impact of time-delayed feedback in two dimensions. 
	The particle dynamics is described by an overdamped Langevin equation with discrete time delay~\cite{kopp_persistent_2023-1},
	\begin{equation}
		\gamma \dot{\bm{r}} = \bm{F}(\bm{r}(t),\bm{r}(t-\tau)) + \bm{\xi}(t), \label{eq:sdde}
	\end{equation}
	where the time-dependent position of the particle is described by $\bm{r} = (x,y)^T=\bm{r}(t)$, $\gamma$ is the friction constant, and $\tau$ is the delay time. 
	Furthermore, $\bm{\xi}(t)$ represents a two-dimensional Gaussian white noise with zero mean and correlation function 
	$\langle \xi_\alpha (t) \xi_\beta (t^\prime) \rangle = 2 \gamma k_B T \delta(t-t^\prime)\delta_{\alpha \beta}$, 
	where $\alpha$, $\beta$ represent the Cartesian components of the vector $\bm{\xi}(t)$ and $k_B T$ (with $k_B$ being the Boltzmann constant and $T$ the temperature) is the thermal energy. 
	The diffusion constant of the free particle motion ($\bm{F}=0$) follows from the Stokes-Einstein relation $D=k_B T/\gamma$.
	The feedback force $\bm{F}$ in Eq.~(\ref{eq:sdde}) is derived from a time-delayed, Gaussian feedback potential, i.e. $\bm{F} = -\nabla_{\bm{r}} U(\bm{r}(t),\bm{r}(t-\tau))$, where
	\begin{equation}
		U(\bm{r}(t),\bm{r}(t-\tau)) = A \exp\left\lbrace - \frac{\left[\bm{r}(t) - \bm{r}(t-\tau)\right]^2}{2b^2}\right\rbrace.
		\label{eq:nonlinear_potential}
	\end{equation}
	The time-delayed force thus depends on both, the position at the current time $t$ and that at the earlier time $t-\tau$.
	%%%%%%%%%%%%%%%%%%%%%%%%%%%%%%%%%%%%%%%%%%%%%%%%%%%%%%%%%%%%%%%%%%%%%%%%%%%%%%%%%%%%%%%%%%%%%%%%%%%%%%%%%%%%%%%%%%%%%%%%%%
	Both the feedback range $b$ (setting the length scale in our system) and the feedback strength $A$ are positive constants. Thus, we consider a repulsive, Gaussian time-delayed feedback potential with 
	a finite range. This time-dependent force acts only on the particle but not on the bath (i.e. the solvent). The numerical solution of Eq.~\eqref{eq:sdde} is described in Appendix~\ref{sec:num_sdde}.
	In a previous study~\cite{kopp_persistent_2023-1} we have shown that the feedback force derived from Eq.~(\ref{eq:nonlinear_potential}) can be interpreted as a source of a ``nudge'' following the particle with some delay. This nudge can generate persistent motion at feedback parameters above a threshold in parameter space. 
	More specifically, feedback parameters fulfilling the condition
	$A\tau/\gamma b^2 > 1$ lead to a constant, long-time velocity vector $\bm{v}_\infty$ of the particle in the deterministic limit due to the nonlinearity of the feedback force \cite{kopp_persistent_2023-1}. The
	magnitude of this velocity can be obtained analytically,
	it is given by $v_\infty = (\sqrt{2}b/\tau) \sqrt{-\ln(\gamma b^2/A\tau)}$~\cite{kopp_persistent_2023-1}. 	
	Below the threshold value $A\tau/\gamma b^2 =1$, one has $v_\infty=0$.
	In the presence of fluctuations, the particle moves, in the long-time limit (and for $A\tau/\gamma b^2>1$), again with a velocity of constant magnitude $v$ on average,
	while the direction of motion randomizes after 
	a persistence time $\tau_\mathrm{p}$~\cite{kopp_persistent_2023-1}. On the other hand, for $A\tau/\gamma b^2 < 1$, the time-delayed feedback only results in enhanced diffusion~\cite{kopp_persistent_2023-1}. 
	With this, the behavior of the feedback-driven particle in the persistent regime (and with noise) 
	is similar \cite{kopp_persistent_2023-1,kopp_spontaneous_2023} to that of an active Brownian particle. This has also been demonstrated in recent experiments~\cite{bell-davies_dynamics_2023}. 
	For what follows, we briefly summarize the relevant parameters and their combinations.
	As discussed above, the parameters $A$, $b$ and $\tau$ are all connected via the dimensionless quantity $A\tau /\gamma b^2$, which fully characterizes the (deterministic) dynamics and also plays an important role in the stochastic case. The feedback range $b$ and the friction parameter $\gamma$ are both constants. Thus, when varying $A\tau/\gamma b^2$, we essentially vary $A$ or $\tau$. 
	In the presence of fluctuations, we non-dimensionalize these quantities by $A/k_B T$ (feedback amplitude versus thermal noise) and $\tau/\tau_B$ (delay time normalized by the Brownian relaxation time scale), respectively.
		
\begin{figure}[h!]
		\centering
		\includegraphics[width=.425\textwidth]{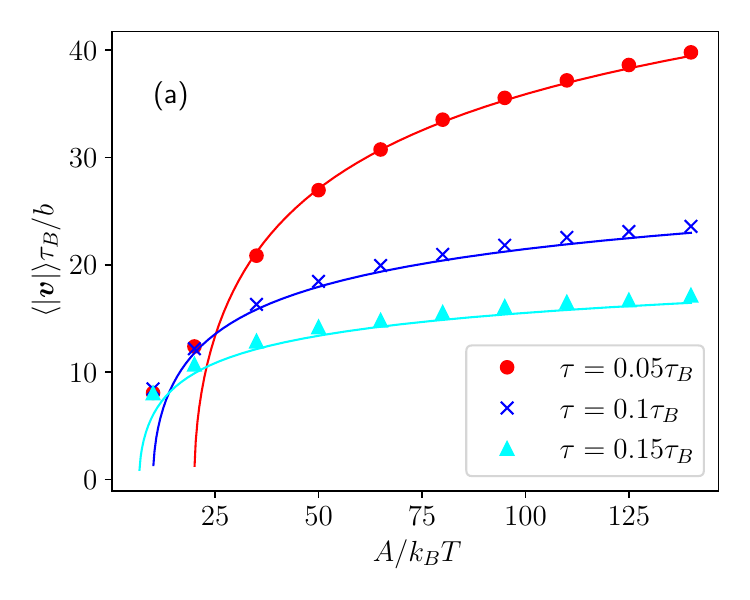}
		\includegraphics[width=.425\textwidth]{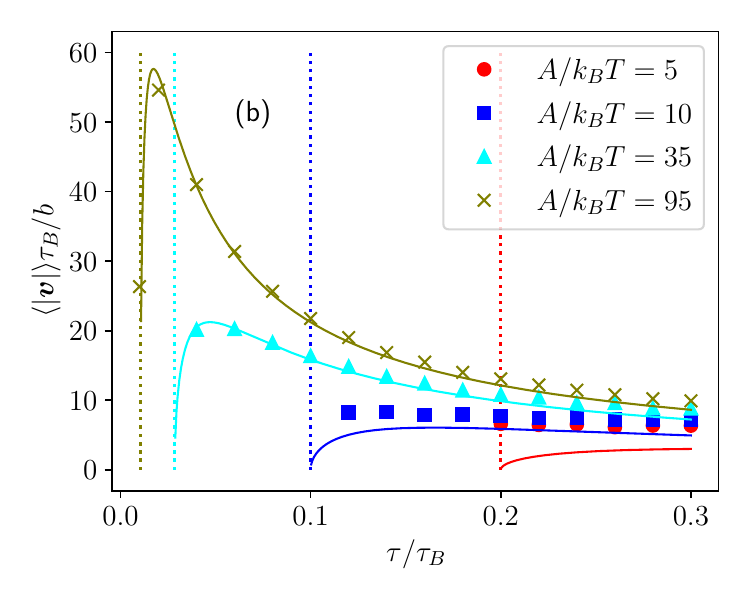}
		\caption{Long-time velocity magnitude $v=\langle|\bm{v}|\rangle$ based on the numerical simulation of Eq.~\eqref{eq:sdde} (symbols) [with feedback force derived 
				from Eq.~\eqref{eq:nonlinear_potential}], and deterministic value $v_\infty$ (solid lines). (a) Results as a function of the dimensionless feedback strength $A/k_BT$ 
				for three different values of the delay time $\tau$ (color-coded, BD data represented as symbols). (b) Results as a function of the delay time $\tau$ for four different values of $A/k_B T$ (color-coded, BD data represented by symbols). Dotted vertical lines: deterministic persistence threshold (color-coded). 
		}\label{fig:vinfty_vs_A}
	\end{figure}

	As a reference for the following investigations, we plot in Fig.~\ref{fig:vinfty_vs_A} the mean (i.e., noise-averaged) long-time velocity magnitude, $v = |\bm{v}|$, obtained from numerical simulations (symbols) of the stochastic system. We note that the numerical calculation of $v$ is not straightforward in an overdamped system, since $\dot{\bm{r}}$ is directly affected by the noise.
	Here, we calculate $v$ in the regime of persistent motion as detailed in Appendix~\ref{sec:num_sdde}. Also shown in
	Fig.~\ref{fig:vinfty_vs_A} are analytical results for the deterministic long-time 
	velocity $v_\infty$ (solid lines).
	
	In Fig.~\ref{fig:vinfty_vs_A}(a), we focus on the effect of the feedback strength $A$ (in units of $k_B T$) at different values of the delay time $\tau$ (color-coded, BD data represented as symbols) in the "supercritical" regime, $A\tau/\gamma b^2 > 1$. 
	We see that $v$ increases monotonically when increasing the strength of the feedback potential, with the rate of increase being strongest close to the threshold. Upon decreasing $\tau$, the threshold to persistent motion moves towards 
	larger $A$, and at the same time, higher long-time velocities become possible.
	On the other hand, when considering $v$ as a function of the delay time [Fig.~\ref{fig:vinfty_vs_A}(b)], we observe a sharp increase of $v$ up to a maximum close above the deterministic persistence threshold for strong enough feedback [two upper curves in Fig.~\ref{fig:vinfty_vs_A}(b)]. 
	After reaching its maximum (for large enough $A$), $v$ slowly decreases. 
		The delay time at the maximum (computed using the expression for the deterministic long-time velocity magnitude $v_\infty$ as a function of $\tau$) reads $\tau_\mathrm{max} = b^2 \sqrt{e} \gamma / A$. 
		This value related to the maximum of $v_\infty$ can be directly related to $v_\mathrm{max}$, the maximum velocity accessible in the deterministic system \cite{kopp_persistent_2023-1}, specifically, $\tau_\mathrm{max} = b/v_\mathrm{max}$, where $v_\infty(\tau_\mathrm{max}) = v_\mathrm{max}= A \exp(-1/2)/\gamma b$. We note that the very presence of a maximum velocity and, thus, a maximum feedback force, is a consequence of the nonlinearity of the system \cite{kopp_persistent_2023-1}: the maximum feedback force, $|\bm{F}_\mathrm{max}| = \exp(-1/2)A/b$, occurs when the displacement within one delay time reaches the range $b$.
		The overall behavior of the function $v_\infty(\tau)$ is then plausible. For $\tau \to 0$, the feedback force vanishes, and, thus, $v_\infty \to 0$. On the other hand, for $\tau \gg \tau_\mathrm{max}$, the feedback force becomes progressively smaller with increasing $\tau$ (yet remains finite for reasonable $\tau$), as a consequence of the growth of the displacement (i.e., $v_\infty \tau$) and the associated exponential `damping'.
		The curves for different $A$ converge at long delay times. All trends seen in Fig.~\ref{fig:vinfty_vs_A} conform with those of the deterministic velocity. Typically, we find $v\gtrsim v_\infty$.
	\subsection{Heat production\label{sec:thermo_quant}}
	We now turn to thermodynamic quantities~\cite{sekimoto_langevin_1998,seifert_stochastic_2012-1}, particularly the rate of heat production.
	Following Sekimoto~\cite{sekimoto_langevin_1998}, we define the increment of heat dissipated into the surrounding heat bath for a given stochastic trajectory as
	\begin{equation}
		\delta q = \left[\gamma\frac{\partial \bm{r}}{\partial t} - \bm{\xi}(t)\right] \circ d\bm{r},
		\label{eq:heatdifferential}
	\end{equation}
	where $\circ$ denotes a Stratonovich product~\cite{gardinerStochasticMethodsHandbook2009}. Equation~\eqref{eq:heatdifferential}
	expresses the fact that heat may be dissipated into the environment due to friction and random forces. From Eq.~\eqref{eq:heatdifferential} we obtain the 
	heat production rate as 
	\begin{equation}
		\dot{q} = \frac{\delta q}{dt} = \left[\gamma\dot{\bm{r}} - \bm{\xi}(t)\right] \circ \dot{\bm{r}}.
		\label{eq:qdot}
	\end{equation}
	We note that, formally, $\dot{q}$ is closely related to the so-called medium entropy production rate $\dot{s}_m = \dot{q}/T$ \cite{seifert_stochastic_2012-1}. 
		Here, we will primarily discuss the noise-averaged heat production rate, $\langle \dot{q} \rangle$, which may be calculated numerically as
	\begin{equation}
		\langle \dot{q} \rangle = \frac{1}{N}\sum_{n=1}^{N} \left[\gamma\frac{\partial \bm{r}_n}{\partial t} - \bm{\xi_n}(t)\right] \circ \frac{d\bm{r}_n}{dt},\label{eq:num_heat_rate}
	\end{equation}
	where $N$ is the number of noise realizations. Details concerning the Stratonovich calculus in the numerical calculations are described in Appendix~\ref{sec:num_strat}.
	The noise average is taken in the long-time limit, when the system has typically reached a (nonequilibrium) steady state.
	It has been shown that time-delayed feedback alone can lead to a finite steady-state heat production rate~\cite{munakata_linear_2009} and even negative heat flow~\cite{loosMediumEntropyReduction2021}. 
	
	Besides the noise average, we also present results for the distribution $P(q(t))$ of the integrated heat
	\begin{equation}
		q(t) = \int\limits_{0}^{t} \dot{q}(t^\prime) dt^\prime.
		\label{eq:heatintegrated}
	\end{equation}
	In particular, we calculate the variance of $P(q)$, $ \sigma_q^2=\langle (q - \langle q \rangle)^2\rangle$,
	as well as the skewness, $ \langle (q - \langle q \rangle)^3\rangle/\sigma_q^3$. Note that the moments are time-dependent, as is $P(q)$.
	\section{Analytical considerations\label{sec:analytical_considerations}}
	Starting from the definition of the heat production rate~\eqref{eq:qdot} and reinserting the right-hand side of the Langevin equation~\eqref{eq:sdde} we obtain, after averaging over noise,
	\begin{align}
		\langle\dot{q}\rangle &= \langle\bm{F}\left(\bm{r}(t), \bm{r}(t-\tau)\right) \circ \dot{\bm{r}}(t)\rangle \nonumber\\
		&= \langle\bm{F}\left(\bm{r}(t), \bm{r}(t-\tau)\right) \circ \frac{1}{\gamma} \left[\bm{F}\left(\bm{r}(t), \bm{r}(t-\tau)\right) + \bm{\xi}(t)\right]\rangle.\label{eq:heat_rate_analytical}
	\end{align}
	While for linear time-delayed feedback forces, this quantity can be calculated analytically~\cite{munakata_linear_2009,loos_heat_2019}, this is essentially impossible for the nonlinear force 
	derived from Eq.~\eqref{eq:nonlinear_potential}. The reason is that when expanding the exponential in Eq.~\eqref{eq:nonlinear_potential}, generally, all moments of the displacement $\bm{r}(t) - \bm{r}(t-\tau)$ within one delay time and, thus, all 
	spatiotemporal correlations of the non-Markovian system, would be needed. Therefore, in the nonlinear case, no exact expressions for  $\langle\dot{q}\rangle$ can be derived.
	An exception is the case of a constant driving force (which is an approximation for the regime of persistent motion) that we discuss in Sec.~\ref{sec:persistent_motion}.
	
	Here, we first focus on situations where the feedback force can be linearized.
	\subsection{\label{sec:linear_case}Linear case}
	We first consider the case of small displacements within one delay time. This corresponds to small feedback strength $A/k_B T$ and/or small values of the delay time $\tau$. 
	In this situation, we can linearize the feedback force~\cite{kopp_persistent_2023-1}, i.e. $\bm{F} \approx (A/b^2) \left[\bm{r}(t) - \bm{r}(t-\tau)\right]$, resulting
	in the equation of motion
	\begin{equation}
		\gamma \dot{\bm{r}} = \frac{A}{b^2} \left[\bm{r}(t) - \bm{r}(t-\tau)\right] + \bm{\xi}(t).\label{eq:sdde_linear}
	\end{equation}
	The average heat production rate then reduces to 
	\begin{align}
		\langle \dot{q} \rangle &= \frac{1}{\gamma} \left(\frac{A}{b^2}\right)^2\langle |\bm{r}(t) - \bm{r}(t-\tau)|^2 \rangle \nonumber \\
		&~~+ \frac{A}{b^2} 2D - \frac{A}{b^2} 2D \delta_{\tau} \label{eq:heat_rate_displacement}
	\end{align}
	(see Appendix~\ref{sec:linear_feedback_force} for details on the derivation).
	For $\tau > 0$, there are two positive terms on the right-hand side of Eq.~\eqref{eq:heat_rate_displacement}. Therefore, $\langle\dot{q}\rangle$ is nonzero and positive for any finite delay time. On the other hand, directly at $\tau=0$, the displacement term is zero since $\bm{r}(t) = \bm{r}(t-\tau)$, and the remaining terms cancel each other. Thus, $\langle \dot{q}\rangle$ vanishes at $\tau=0$, as expected for a passive system. These findings imply a discontinuity in the limit $\tau\to 0$ that is a consequence of the overdamped character of the equation of motion \cite{loos_heat_2019}.
	
	We are now interested in the steady-state behavior at $\tau>0$. Yet, here we encounter a problem: Strictly speaking, the linear system in Eq.~\eqref{eq:sdde_linear} has no well-defined stationary solution since the prefactors of $\bm{r}(t)$ and $\bm{r}(t-\tau)$ are the same.
	Indeed, linear stochastic systems with time delay have been extensively studied in the literature~\cite{kuchler_langevins_1992,frank_stationary_2001-2}, and it is useful to recapitulate some results. Let us therefore consider the linear SDDE 
	\begin{equation}
		\gamma \dot{\bm{r}} = k_1\bm{r}(t) - k_2\bm{r}(t-\tau) + \bm{\xi}(t),
		\label{eq:general_linear}
	\end{equation}
	where $k_1$ and $k_2$ are independent coefficients. The equations of motion for $x$- and $y$-direction decouple, allowing for a 
	straightforward generalization to a two-dimensional system from the one-dimensional problems studied in the literature~\cite{kuchler_langevins_1992,frank_stationary_2001-2}.
	As demonstrated in~\cite{kuchler_langevins_1992}, a stationary solution (corresponding to a finite long-time correlation function $\langle\bm{r}(t)\cdot\bm{r}(t-\tau)\rangle$) exists if either $k_1-k_2<0$ and $k_1+k_2\leq 0$, or if $k_1-k_2<0$ and $k_1+k_2>0$ for $\tau \in \left(0,r_0(k_1,k_2)\right)$, with 
	$r_0(k_1,k_2) = \left[\arccos(k_1/k_2)/(k_2^2 - k_1^2)^{1/2}\right]$. In our situation, $k_1 = k_2 = A/b^2$ [see Eqs.~\eqref{eq:sdde_linear} and~\eqref{eq:general_linear}], i.e. we are just on the boundary of the stable region. 
	This corresponds to a marginally stable situation.
	Below we first consider $\langle\dot{q}\rangle$ in a truly stable situation, specifically $k_1<k_2$. This will allow us, at the same time, to benchmark our numerical simulations, and to explore the behavior of $\langle\dot{q}\rangle$ when approaching the boundary $k_1\to k_2$.
	
	\subsubsection{Heat production rate in stable linear sytems\label{sec:stable_system}}
\begin{figure}
		\centering
		\includegraphics[width=0.425\textwidth]{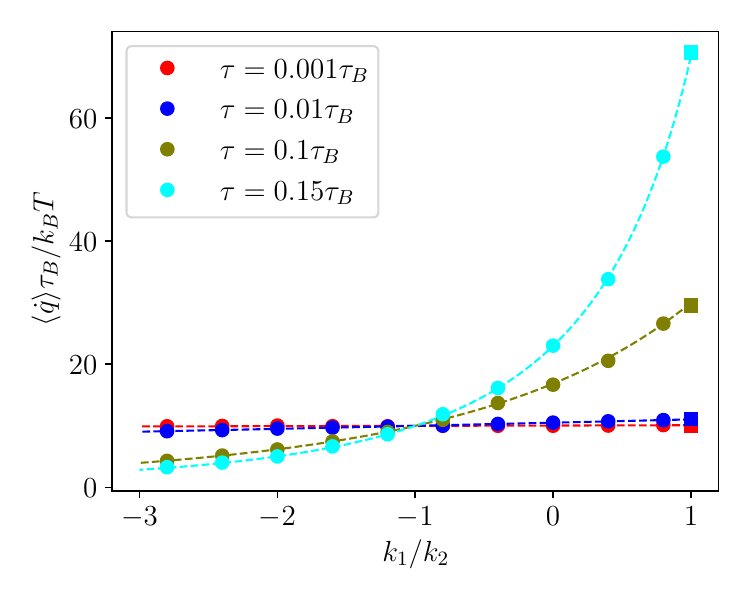}
		\caption{Long-time, noise-averaged heat production rate plotted as function of the ratio of prefactors $k_1$ and $k_2$ in Eq.~(\ref{eq:general_linear}), at different delay times $\tau$ (color-coded). 
				Circles: BD data in the stable regime $k_1/k_2<1$ at different $\tau$. Squares: BD data in the limit $k_1 = k_2$. Dashed lines: analytical results for stable systems [see Eq.~\eqref{eq:analytical_expression_heat_rate}]. 
				Parameters: $k_2 = A/b^2 = 5 k_B T/b^2$ (i.e., below the deterministic persistence threshold for all delay times $\tau$ considered here).}\label{fig:analyticalvsnumerical}
	\end{figure}

	Choosing $k_1$ and $k_2$ such that a stationary state exists, $\langle \dot{q}\rangle$ can be calculated using known results for the (one-dimensional analogs of the) two-point correlation functions $\langle\bm{r}(t)\cdot\bm{r}(t-\tau)\rangle$ and $\langle\bm{r}(t)^2\rangle = \langle\bm{r}(t-\tau)^2\rangle$~\cite{kuchler_langevins_1992}, as well as for the position-noise correlations $\langle\bm{\xi}(t)\cdot\bm{r}(t)\rangle$ and $\langle\bm{\xi}(t)\cdot\bm{r}(t-\tau)\rangle$~\cite{frank_stationary_2001-2,loos_heat_2019}.
	Since in Eq.~\eqref{eq:general_linear}, the equations for the $x$- and $y$-components decouple, $\langle\dot{q}\rangle$ is given by the sum of the contributions from the $x$- and $y$-motion, for which an analytical expression is known~\cite{munakata_linear_2009,loos_heat_2019}. The steady-state result then reads
	\begin{equation}
		\frac{\langle \dot{q}\rangle_{\mathrm{ss}}}{2D} = -k_1 - k_2\delta_\tau - \left(k_1^2 - k_2^2\right) \frac{1 + \frac{k_2 \sinh \left(\frac{\tau}{\gamma}\sqrt{k_1^2 - k_2^2}\right)}{\sqrt{k_1^2 - k_2^2}}}{-k_1 + k_2\cosh \left(\frac{\tau}{\gamma}\sqrt{k_1^2 - k_2^2}\right)}.\label{eq:analytical_expression_heat_rate}
	\end{equation}
	For vanishing delay times ($\tau = 0$), $\langle \dot{q} \rangle_{\mathrm{ss}} /2D = - k_1 - k_2 + (k_1 + k_2) = 0$, as expected.
	We also see that for finite $\tau$, in the limit $k_1\to -k_2$, $\langle \dot{q} \rangle_{\mathrm{ss}} = 2D A/b^2$, i.e. the heat production rate becomes finite and independent of the (finite) delay time.
	We now utilize Eq.~(\ref{eq:analytical_expression_heat_rate}) with $k_1<k_2$ as a benchmark for our numerical (BD) calculations, see Fig.~\ref{fig:analyticalvsnumerical}.
	The numerical results indeed match those from Eq.~(\ref{eq:analytical_expression_heat_rate}), including the case $-k_1 = k_2$ where $\langle \dot{q}\rangle_{\mathrm{ss}}$ becomes independent 
	of the delay time.
	
	We now consider the limit $k_1 \to k_2$. In this limit, Eq.~(\ref{eq:analytical_expression_heat_rate}) still gives a result. However, this has to be taken with care since, strictly speaking,
	$k_1=k_2$ is just outside the region where a stationary solution of the linear SDDE exists~\cite{kuchler_langevins_1992}. Still, it is interesting to see how the numerical results behave when we approach this limit 
	starting from the stable regime. The results of this "extrapolation" are indicated by squares in Fig.~\ref{fig:analyticalvsnumerical}. Overall, we find that there are no discontinuities or other sudden changes in qualitative behavior of $\langle\dot{q}\rangle$ when approaching the limit $k_1 \to k_2$ from below. 
	
	How can it be that we can (numerically) obtain steady-state results for $k_1=k_2$? The reason lies in our choice of feedback parameters.
	In Fig.~\ref{fig:analyticalvsnumerical} these are chosen such that, in all cases,
	$A\tau/\gamma b^2<1$. Indeed, as we have shown in \cite{kopp_persistent_2023-1}, the case $A\tau/\gamma b^2=1$ corresponds to a critical value not only for the nonlinear feedback (where it is related to the onset of persistent motion), but also 
	in the linear case. Analyzing the deterministic equation one finds that, below this value, the particle comes to rest in the long-time limit, i.e., its displacement goes to zero. For explicit results see Appendix~\ref{sec:finite_displacements}.
	This stability at $A\tau/\gamma b^2<1$ remains in the stochastic case (at least at the noise parameters considered).
	At $A\tau/\gamma b^2>1$, on the other hand, the displacement obtained deterministically grows indefinitely, yielding a divergence of the force.
	\subsubsection{Small delay approximation\label{sec:small_delay_approx}}
	We now turn back to Eq.~\eqref{eq:heat_rate_displacement}, this time focusing on small, yet finite delay times. In this limit, as shown in \cite{kopp_persistent_2023-1}, the particle's motion reduces to that of a \textit{free} Brownian particle with 
	renormalized diffusion coefficient $D_\mathrm{SD} = D \left(1+A\tau/\gamma b^2\right)^2$. Consequently, the displacement within one delay time becomes  $\langle |\bm{r}(t) - \bm{r}(t-\tau)|^2 \rangle = 4D_\mathrm{SD}\tau$.
	Inserting this expression into Eq.~\eqref{eq:heat_rate_displacement} we obtain
	\begin{align}
		\langle \dot{q} \rangle &\approx \frac{1}{\gamma} \left(\frac{A}{b^2}\right)^2\left[4D\left(1+\frac{A\tau}{\gamma b^2}\right)^2\right]\tau \nonumber\\ 
		&~~+ \frac{A}{b^2} 2D - \frac{A}{b^2}2D\delta_\tau.
		\label{eq:MSD_approx}
	\end{align}
	Eq.~\eqref{eq:MSD_approx} reproduces the vanishing of $\langle \dot{q} \rangle$ at $\tau=0$, and the discontinuity upon increasing $\tau$ from zero. 
	Indeed, for infinitesimal $\tau$, the first term on the right side is negligible, such that $\langle \dot{q} \rangle\approx A2D/b^2$.
	We further note that the small-delay approximation also holds for the nonlinear system in the limit of small $\tau$, since small delay leads to an effective linearization~\cite{kopp_persistent_2023-1}.
	\subsection{\label{sec:persistent_motion}Regime of persistent motion}
	We finally turn to the regime of large feedback parameters beyond the deterministic threshold, i.e., $A\tau/\gamma b^2 > 1$. Here, the full nonlinearity of the feedback force becomes relevant, and the particle
	develops persistent motion characterized by a constant long-time velocity magnitude $v$ (after averaging over noise), see Fig.~\ref{fig:vinfty_vs_A}.
	To derive an analytical approximation for $\langle\dot{q}\rangle$ we note that a constant velocity corresponds to a constant force, $F$. For such a system, the steady-state heat production rate can be easily calculated from Eq.~(\ref{eq:heat_rate_analytical}), yielding
	$\langle \dot{q} \rangle=F^2/\gamma=\gamma v^2$. As an analytical estimate, we replace $v$ by its deterministic counterpart, $v_\infty$, yielding
	\begin{equation}
		\langle \dot{q} \rangle_\mathrm{det} = \gamma v_\infty^2.\label{eq:heat_rate_constant}
	\end{equation}
	where the subscript ``det" refers to the deterministic limit.
	
	The limit of a constant force also provides a reference for the distribution of the integrated heat, $P(q(t))$, with $q(t)$ given in Eq.~(\ref{eq:heatintegrated}). 
	A straightforward calculation shows that the distribution is Gaussian, where the mean and variance are given by
	\begin{align}
		\langle q(t)\rangle_\mathrm{det} & = \gamma v_\infty^2t,\label{eq:frist_moment_det}\\
		\left(\langle q^2(t)\rangle-\langle q(t)\rangle^2\right)_\mathrm{det} & = 2\gamma k_B Tv_\infty^2 t\\
		& = 2 k_B T\langle q(t)\rangle_\mathrm{det}.
		\label{eq:moments_det}
	\end{align}
	Thus, the first two moments grow linearly in time (see also, e.g., \cite{gomez-solanoNonequilibriumWorkDistribution2015}). Higher cumulants vanish.
	%
	%%%%%%%%%%%%%%%%%%%%%%%%%%%%%%%%%%%%%%%%%%%%%%%%%%%%%%%%%%%%%%%%%%%%%%%%%%%%%%%%%%%%%%%%%%%%%%%%%%%%%%%%%%%%%%%%%%%%%%%%%%%%%%%%%%%%%%%%%%%%%%%
	%
	%
	%
	%
	%%%%%%%%%%%%%%%%%%%%%%%%%%%%%%%%%%%%%%%%%%%%%%%%%%%%%%%%%%%%%%%%%%%%%%%%%%%%%%%%%%%%%%%%%%%%%%%%%%%%%%%%%%%%%%%%%%%%%%%%%%%%%%%%%%%%%%%%%%%%%%%
	\section{\label{sec:numerical_results}Numerical results}
	\subsection{Average heat production}
	\subsubsection{Small feedback}
\begin{figure}[h!]
		\centering
		\includegraphics[width=.425\textwidth]{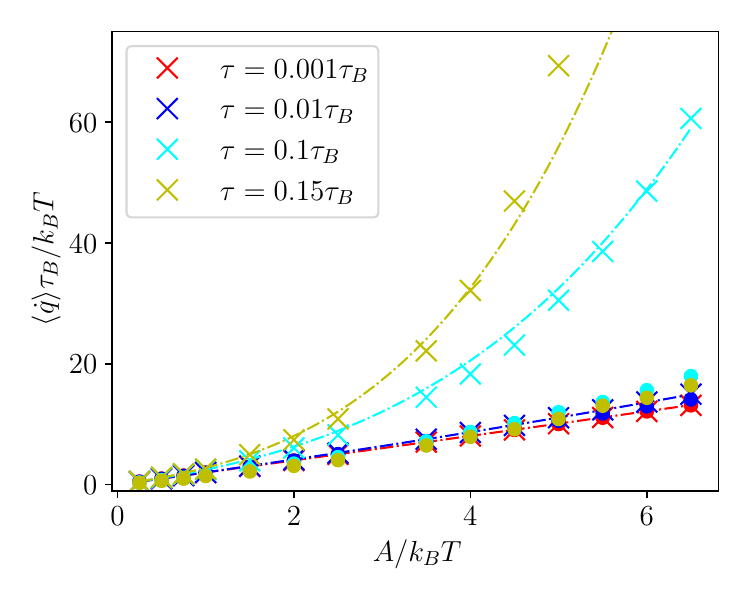}
		\caption{Noise-averaged, long-time heat production rate as a function of $A/k_B T$ for different values of the delay time $\tau$ below the deterministic persistence threshold ($A\tau/\gamma b^2 = 1$).
				Circles: Numerical results based on Eq.~\eqref{eq:sdde} (full, nonlinear feedback force). Crosses: numerical results for the linearized system Eq.~\eqref{eq:sdde_linear} (we have omitted the data points for $A/k_B T > 5$ at $\tau = 0.15 \tau_B$).
				Dashed line: Small-delay approximation Eq.~\eqref{eq:MSD_approx}.}\label{fig:small_displacement_limit}
	\end{figure}

	We first focus on the situation where $A\tau/\gamma b^2 < 1$, i.e., below the threshold of persistent motion in the deterministic limit
	(and at the stability limit of the linearized system). We henceforth refer to this case as the "subcritical" regime.
	In Fig.~\ref{fig:small_displacement_limit} we show the long-time heat production rate for various delay times from numerical simulations (for details on the numerical procedure see Appendix \ref{sec:num_sdde}) based on the SDDE of the full (nonlinear) 
	problem [see Eqs.~\eqref{eq:sdde} and~\eqref{eq:nonlinear_potential}] (circles), as well as corresponding results based on Eq.~\eqref{eq:sdde_linear} for linearized 
	feedback forces (crosses). Also shown are analytical results of the small delay approximation [see Eq.~\eqref{eq:MSD_approx}].
	Note that we have non-dimensionalized the heat production rate $\langle\dot q\rangle$ by using the Brownian time ($\tau_B$) and the thermal energy, $k_B T$. The latter serves as a measure of the noise strength and is assumed to be fixed here.
	
	For both the nonlinear and the linear system, as well as in the limit of small delay, the results in Fig.~\ref{fig:small_displacement_limit} show that the heat production rate is nonzero and positive for $A/k_BT>0$, despite the absence of persistent motion at the parameters considered. 
	We generally observe a monotonic increase with the feedback strength, with the details depending on the type of system (linear versus nonlinear) and the delay time.
	In the linear case, the data (crosses) reveal a strong dependency on $\tau$: the larger the delay, the faster the increase of $\langle\dot q\rangle$ with $A/k_B T$. This trend is well reproduced by the small delay approximation. The latter yields
	results quite close to those in the linear case for all but the largest $\tau$ considered here ($\tau=0.15\tau_B$).
	For the nonlinear problem, the increase of $\langle\dot q\rangle$ with $A/k_B T$ is generally slower than in the linear case, particularly at the two larger delay times ($\tau\geq 0.1\tau_B$) considered in 
	Fig.~\ref{fig:small_displacement_limit}. To understand these differences, we note that in the nonlinear case, the (squared) force entering  $\langle\dot q\rangle$ contains an exponential (Gaussian) factor 
	that stems from the potential [Eq.~(\ref{eq:nonlinear_potential})] and leads to a lowering relative to the linearized force. Also, the dependency 
	on $\tau$ at fixed $A/k_B T$ is much weaker in the nonlinear case. Closer inspection shows that, for fixed $A/k_B T$, the impact of $\tau$ (for long $\tau$) is even opposite to the linear case, i.e., 
	$\langle\dot q\rangle$ decreases with $\tau$. Overall, the data in Fig.~\ref{fig:small_displacement_limit} suggest that the linear approximation of the problem works well at small feedback strength and small delay times, as expected.

\begin{figure}
		\centering
		\includegraphics[width = 0.425\textwidth]{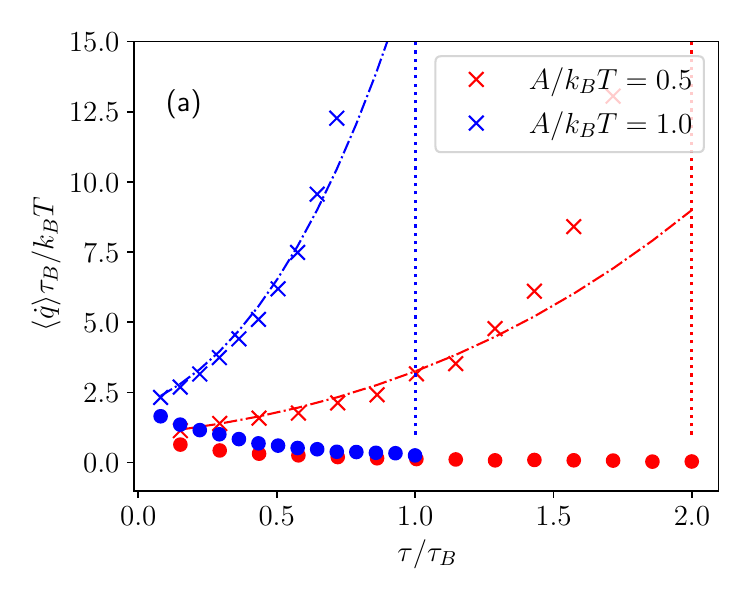}		
		\includegraphics[width = 0.425\textwidth]{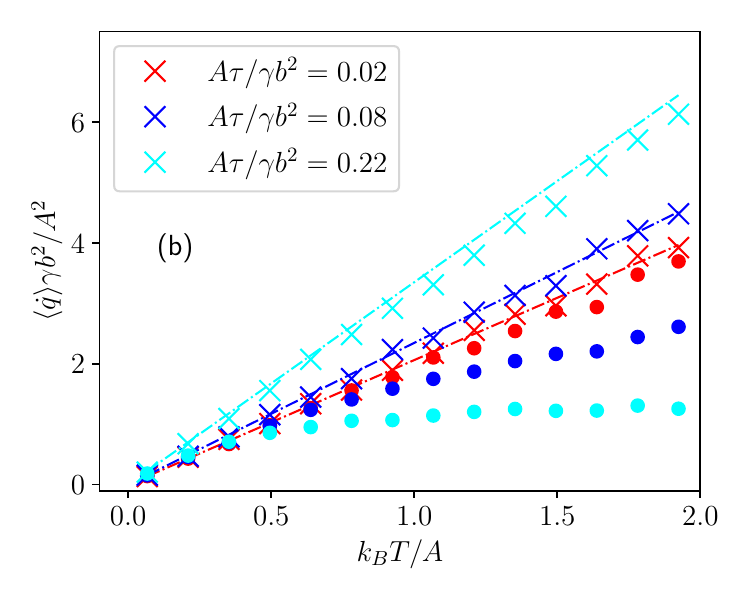}
		\caption{Heat production rate from numerical calculations for the linear (crosses) and nonlinear (circles) system, as well as from the small-delay approximation [see Eq.~\eqref{eq:MSD_approx}] (dashdotted lines) at parameters (color-coded) below the threshold ($A\tau/\gamma b^2 = 1$).
				(a) Results as function of dimensionless delay time (with vertical lines indicating the threshold); (b) Results as function of the relative noise strength.} \label{fig:heat_rate_linear_vs_kBT_tau}
	\end{figure}

	The role of the delay time is further illustrated in 
	Fig.~\ref{fig:heat_rate_linear_vs_kBT_tau}(a) where we plot $\langle\dot q\rangle$ directly as function of $\tau/\tau_B$. When $\tau$ tends to zero, the numerical data for both, the linear and the nonlinear system tend to the same {\em finite} value. 
	(recall that $\langle \dot q\rangle$ vanishes only when $\tau=0$). This finite value is given by $A2D/b^2$, as seen from Eq.~(\ref{eq:MSD_approx}). 
	Upon increasing $\tau$ from small values, the heat production rate progressively increases and even diverges when approaching the threshold, as expected from the divergence of displacements discussed in Sec.~\ref{sec:linear_case}.
	The small-delay approximation first follows the behavior of the linear system but does not reflect the instability. 
	Considering now the full, nonlinear system we see only a weak $\tau$-dependency in the regime considered, consistent with Fig.~\ref{fig:small_displacement_limit}.
	
	Finally, it is interesting to note that the behavior of the linear and nonlinear system differ also with respect to the noise strength.
	To see this, we plot in Fig.~\ref{fig:heat_rate_linear_vs_kBT_tau}(b) the heat production rate scaled by the (now fixed) parameters $A$, $b$, and $\gamma$ as function of $k_BT$. The latter is linearly related to the diffusion constant
	and, thus, serves as a measure of noise in Eq.~(\ref{eq:sdde}). We choose several values of the delay time, corresponding to different values of $A\tau/\gamma b^2<1$.
	It is seen that the linear system shows an essentially linear growth with the noise, which conforms with the small-delay approximation Eq.~\eqref{eq:MSD_approx}. This growth is weakened in the nonlinear case and even seems to saturate
	when the parameter $A\tau/\gamma b^2$ becomes larger. 
		To explain this behavior, we consider the nonlinear feedback force which is directly connected to the heat production (see Eq.~\eqref{eq:heat_rate_analytical}). As explained in Sec.~II A, the (nonlinear) feedback force has a maximum when the displacement equals the feedback range $b$. If the displacement exceeds $b$, the feedback force is `damped' by the exponential factor. Such an increase of the displacement can occur, first, by strong noise, that is, in the regime $k_B T \gtrsim A$. Yet, since smaller displacements are still possible (albeit less likely), $\langle \dot{q} \rangle$ does not vanish entirely. The second mechanism leading to larger displacements (and thus, weakening of the force) is an increase of the delay time. This explains the decrease of $\langle \dot{q} \rangle$ with increasing $A\tau/\gamma b^2$ at fixed noise strength (see Fig.~\ref{fig:heat_rate_linear_vs_kBT_tau}(b)). Overall, the results in Fig.~\ref{fig:heat_rate_linear_vs_kBT_tau}(b) for the nonlinear case reveal a subtle interplay of noise and delay, in stark contrast to the behavior of the linear system.
	\subsubsection{Onset of persistent motion and beyond}
\begin{figure}
		\centering
		\includegraphics[width=.425\textwidth]{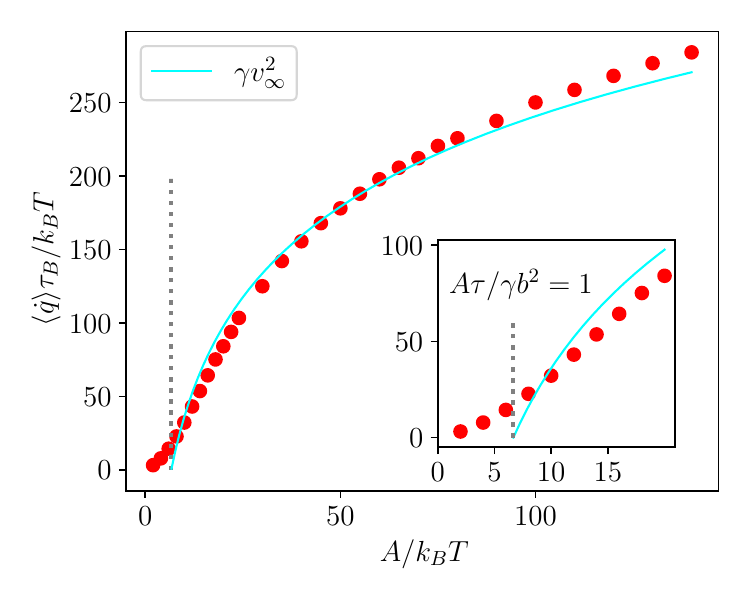}
		\caption{
			Heat production rate for nonlinear time-delayed feedback ($\tau = 0.15 \tau_B$) as function 
				of the feedback strength. Circles: numerical data, solid line: analytical result Eq.~\eqref{eq:heat_rate_constant} for a particle moving with constant velocity $v_\infty$. 
				The vertical dotted line indicates the threshold of persistent motion, $A\tau/\gamma b^2 = 1$; 
				Inset: Zoom-in for small feedback strengths.}\label{fig:heat_rate_function_of_A}
	\end{figure}

	We now consider the heat production rate at parameters where the particle subject to nonlinear feedback develops persistent motion, i.e., a finite velocity, see Fig.~\ref{fig:vinfty_vs_A}.
	To give an overview, we plot in Fig.~\ref{fig:heat_rate_function_of_A} $\langle\dot{q}\rangle$ for a large range of values of $A/k_B T$ from the "sub"- to the "supercritical" regime.
	It is seen that $\langle\dot q\rangle$ behaves smoothly when crossing the threshold of the deterministic case (vertical line): there is no obvious signature besides a subtle change in curvature.
	For the supercritical regime, we have included in Fig.~\ref{fig:heat_rate_function_of_A} analytical results for a system moving with constant velocity $v_{\infty}$, 
	see Eq.~\eqref{eq:heat_rate_constant}. We find that this quasi-deterministic 
	approximation describes the overall trend in the supercritical regime quite well. From a quantitative point of view, the approximation slightly overestimates the values close above the threshold, while the values at large feedback are somewhat too small. We further explore this issue based on an expansion of the deterministic force in Appendix~\ref{sec:corrections}.

\begin{figure}
		\centering
		\includegraphics[width = 0.425\textwidth]{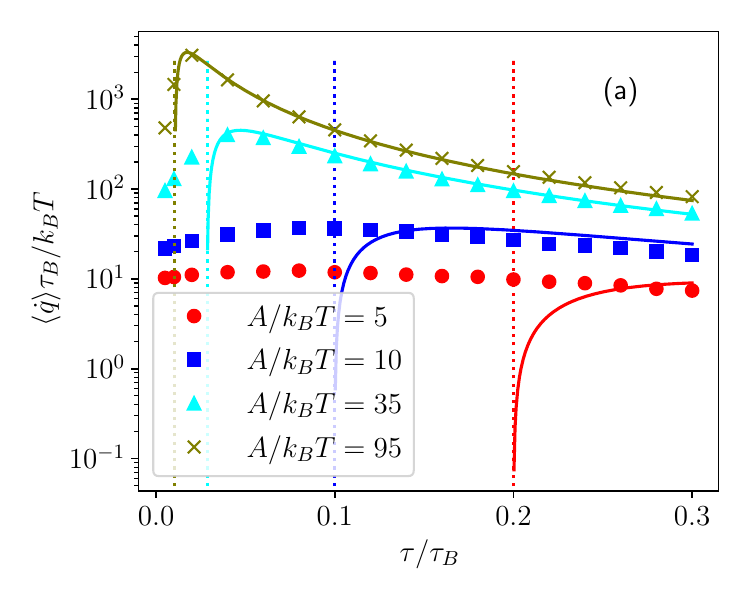}	
		\includegraphics[width = 0.425\textwidth]{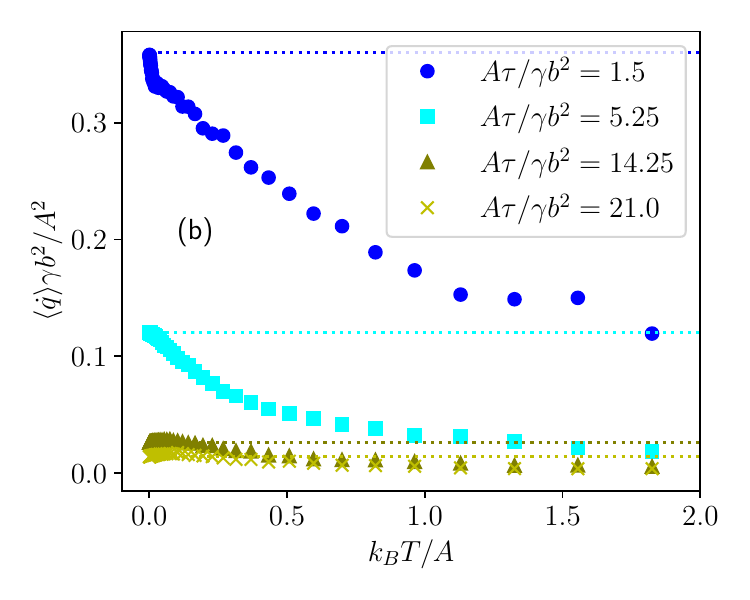}
		\caption{Heat production rate from numerical calculations (symbols) for the nonlinear feedback system at parameters above the threshold, $A\tau/\gamma b^2 > 1$. 
				(a) Results as function of dimensionless delay time. Solid lines refer to the deterministic limit [see Eq.\eqref{eq:heat_rate_constant}]. Vertical dotted lines indicate the $\tau$-values related to the threshold. (b) Results as function of the relative noise strength. Dotted horizontal lines indicate the deterministic limit [see Eq.~(\ref{eq:heat_rate_constant})].} \label{fig:heat_rate_nonlinear_vs_kBT_tau}
	\end{figure}

	Another interesting aspect is the dependency on the delay time that is plotted in Fig.~\ref{fig:heat_rate_nonlinear_vs_kBT_tau}(a). 
		We recall that, for small feedback strength, the heat production rate of the nonlinear system is nearly independent of $\tau$, see Fig.~\ref{fig:heat_rate_linear_vs_kBT_tau}(a).
		The behavior above the threshold is strikingly different. For large values of $A/k_B T$ (such as $A/k_BT = 35$ and $95$), we observe a strong dependency of $\langle\dot q\rangle$ on the delay and even a maximum. This maximum is located close above the deterministic threshold, $A\tau/\gamma b^2 = 1$ (dotted vertical lines, color-coded). A similar observation has been made in Ref.~\cite{bell-davies_dynamics_2023}.

	It is also seen from Fig.~\ref{fig:heat_rate_nonlinear_vs_kBT_tau}(a) that, at large feedback, the behavior is very well described by the quasi-deterministic approximation, Eq.~\eqref{eq:heat_rate_constant}.
	This already suggests that the maximum is related to that of the (deterministic) velocity $v_\infty$ as function of delay, see Fig.~\ref{fig:vinfty_vs_A}(b). 
			We recall (see Sec.~\ref{sec:twoA}) that the maximum of $v_\infty(\tau)$ arises due to the nonlinearity of the feedback force which becomes maximal at $\tau = \tau_\mathrm{max}$. Our numerical results in Fig.~\ref{fig:heat_rate_nonlinear_vs_kBT_tau}(a) show that this behavior of the driving force is directly translated to that of the heat dissipated into the medium. On the one hand, this observation indicates a tight connection between the diving feedback force and the heat dissipated into the surrounding bath, and, on the other hand, reflects the effect of the nonlinearity of the feedback force (which causes the maximum, as discussed in Sec.~\ref{sec:twoA}) on both, $v$ and $\dot{q}$.
		However, the analogy to the deterministic case breaks down when we consider values of $\tau$ corresponding to the subcritical regime, where the deterministic velocity is zero. Here, the numerical data for $\langle\dot{q}\rangle$
	reveal nearly constant behavior. This is consistent
	with what we saw at small feedback strength (Fig.~\ref{fig:heat_rate_linear_vs_kBT_tau}), where the nonlinear and linear system's behavior merges (and both show a discontinuity in the limit $\tau\to 0$).
	
\begin{figure}
		\centering
		\includegraphics[width = 0.425\textwidth]{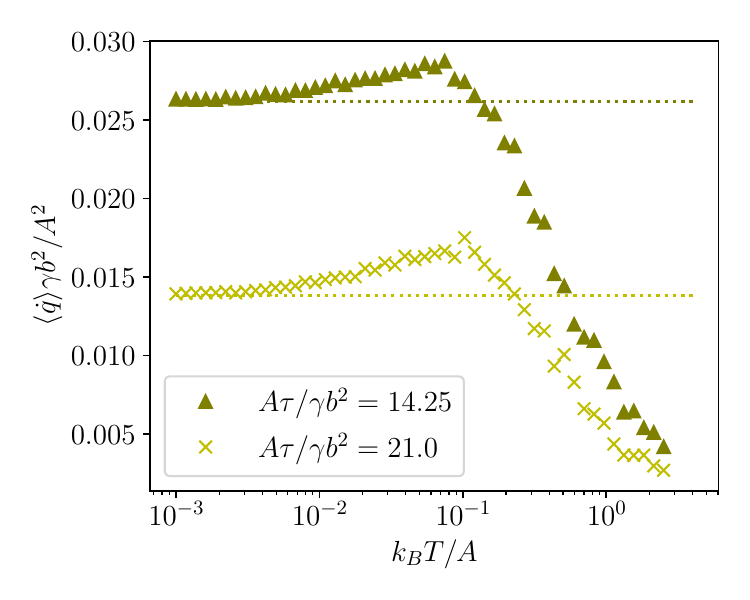}
		\caption{Enlarged version of Fig.~\ref{fig:heat_rate_nonlinear_vs_kBT_tau} for the two largest values of $A\tau/\gamma b^2$ in a semi-logarithmic representation.}\label{fig:heat_rate_nonlinear_vs_kBT_tau_2}
	\end{figure}

	We now turn to the impact of noise (measured by $k_B T/A$ with $A$ fixed).
	In Fig.~\ref{fig:heat_rate_nonlinear_vs_kBT_tau}(b) as well as in Fig.~\ref{fig:heat_rate_nonlinear_vs_kBT_tau_2} we show results for several values of $A\tau/\gamma b^2$ above the threshold.
	We first note that, in all cases, the limit of zero noise leads to finite values of the scaled heat production rate. This limit is given by the (analytically derived) quasi-deterministic value $\gamma v_\infty^2$ [see Eq.~(\ref{eq:heat_rate_constant})], with $v_\infty$ being nonzero at the parameters considered. Upon increasing the noise strength from zero we observe two types of behaviors. For $A\tau/\gamma b^2=1.5$ and $5.25$, the heat production rate decreases monotonically
	with $k_BT/A$, similar to the large-noise behavior seen in the subcritical regime (see Fig.~\ref{fig:heat_rate_linear_vs_kBT_tau}(b)).
	In the here considered supercritical regime, we relate the noise-induced decrease of the heat production rate to growing perturbations of the constant velocity state of the deterministic system. In other words, the effective velocity of the stochastic system becomes successively smaller with noise.
	Contrary to this monotonic decrease, for even larger  $A\tau/\gamma b^2$, we see in Fig.~\ref{fig:heat_rate_nonlinear_vs_kBT_tau}(b) and particularly Fig.~\ref{fig:heat_rate_nonlinear_vs_kBT_tau_2}
	that only after a local maximum where $k_BT/A \approx 0.1$, the heat production rate starts
	to decrease. The two regimes where $\langle \dot q\rangle$ grows and decreases, respectively, correspond to the over- and underestimation of the true heat production rate by the deterministic 
	data in Fig.~\ref{fig:heat_rate_function_of_A}.
	\subsection{Distribution functions\label{sec:distribution}}
	So far we have focused on the \textit{mean} of the (fluctuating) heat production rate. In this last section, we briefly discuss some aspects of the distribution 
	$P(q(t))$ of the (fluctuating) heat dissipated over some time interval of length $t$, see Eq.~(\ref{eq:heatintegrated}).
	To our knowledge, there are only very few literature results for thermodynamic distribution functions of time-delayed systems, and these refer to specific linear systems differing from those considered here
	\cite{rosinbergFluctuationsDynamicalObservables2024,loosMediumEntropyReduction2021}. Therefore, there are no ``benchmark" results.
	
\begin{figure*}
		\centering
		\includegraphics[width=0.9\textwidth]{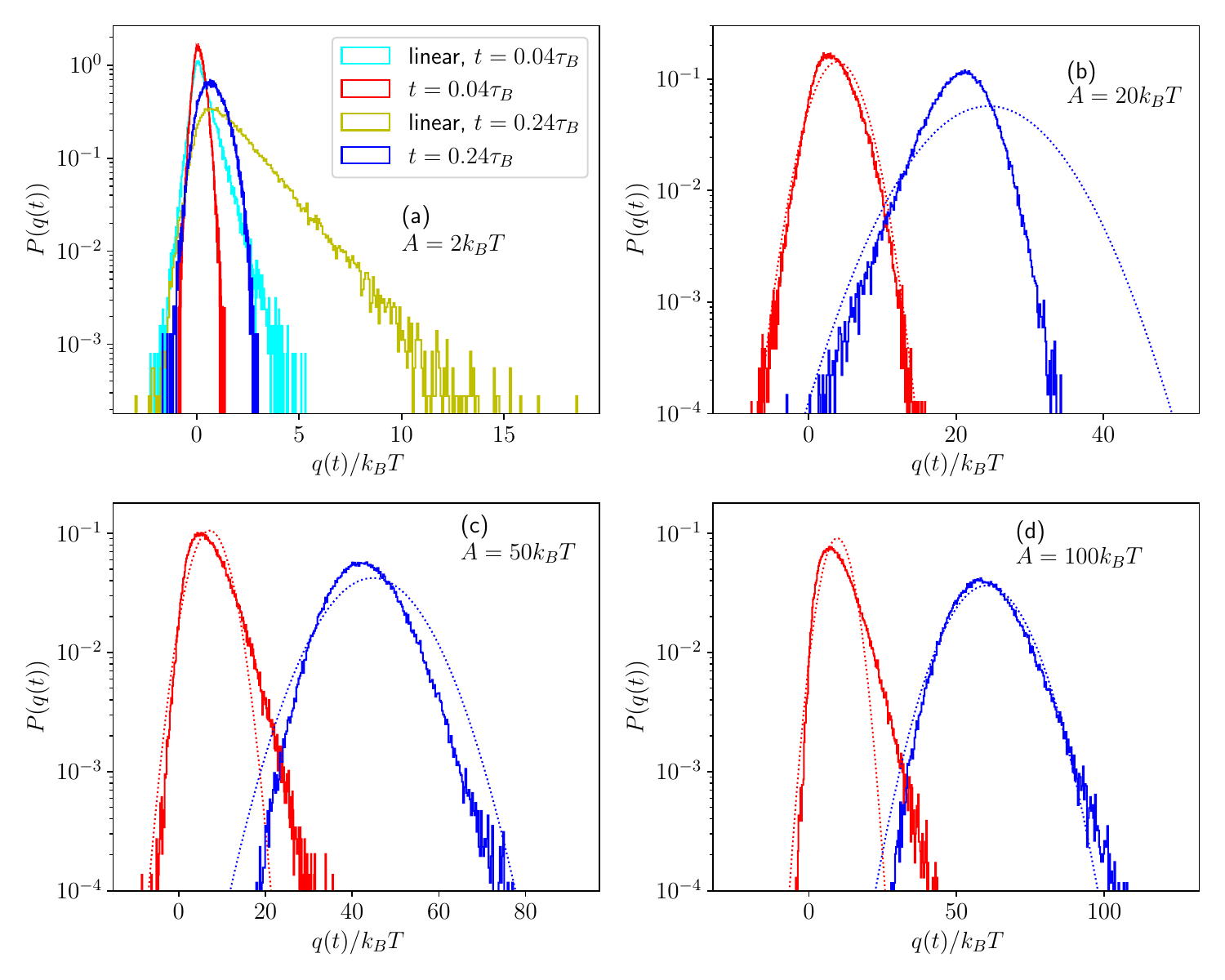}
		\caption{Numerical results for the distribution of the dissipated heat, $P(q(t))$, based on Eq.~\eqref{eq:sdde} (nonlinear feedback). Panels (a)-(d) refer to different values of $A/k_BT$, each containing results for two integration times (color-coded according to the legend in (a)). Panel (a) also shows results for the corresponding linear systems. In (b)-(d), dashed lines correspond to the Gaussian distribution of a particle subject to a constant force [Eqs.~\eqref{eq:frist_moment_det} - (\ref{eq:moments_det})]. In all cases, $\tau = 0.15 \tau_B$.
		}\label{fig:distributions_heat_tau_015}
	\end{figure*}

\begin{figure}[h!]
		\centering
		\includegraphics[width=0.425\textwidth]{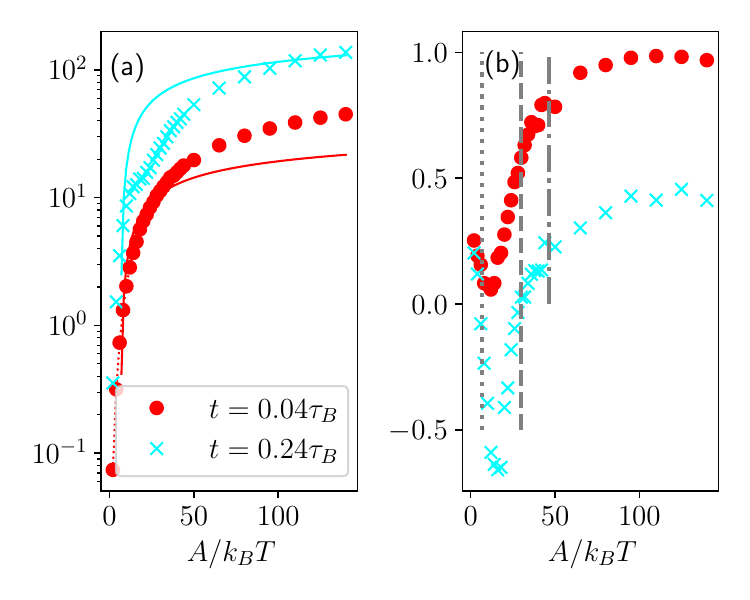}
		\caption{Numerical results for (a) the variance and (b) the third normalized cumulant (skewness) of the distribution $P(q(t))$ of the nonlinear system as functions of feedback strengths and the two integration times considered in Fig.~\ref{fig:distributions_heat_tau_015}. The solid lines in (a) correspond to the deterministic limit of the variance, see Eq.~(\ref{eq:moments_det}).
				In panel (b), the vertical dotted line indicates the deterministic persistence threshold. The other vertical lines are explained in the main text. \label{fig:moments_tau_015}}
	\end{figure}

	In the four panels of Fig.~\ref{fig:distributions_heat_tau_015}, we plot numerical results for $P(q)$ of the nonlinear system below (panel (a)) and above (panels (b)-(d)) the threshold to persistent motion. 
	We note that, by definition, $P(q)$ strongly depends on $t$. In particular, the mean increases linearly with $t$ (this constant slope is expected in view of our steady-state results for $\langle\dot{q}\rangle$, and we have checked this explicitly). In Fig.~\ref{fig:distributions_heat_tau_015},
	we consider two integration times, one being smaller, the other being larger than the delay time.
	In the subcritical regime [Fig.~\ref{fig:distributions_heat_tau_015}(a)], the distributions appear to be fairly symmetric at both integration times considered.
	The width increases with time, but the distributions remain much narrower than those at the same integration time and larger feedback strength (see panels (b)-(d)). 
	This is more directly seen in Fig.~\ref{fig:moments_tau_015}(a) where we plot the variance of the distribution (for the two integration times) as function of $A/k_BT$.
	We interpret the relatively small variances in the subcritical regime as a consequence of an effective confinement of the motion, consistent with the small values of the mean heat production rate 
	in this regime (see Fig.~\ref{fig:small_displacement_limit}). Figure~\ref{fig:distributions_heat_tau_015}(a) further shows that the distributions in the subcritical case differ markedly from that of corresponding linear systems.
	The latter are not only broader, but also asymmetric. Moreover, they have exponential tails. Similar behavior has been seen in the distribution of entropy production of linear systems with distributed delay \cite{loosMediumEntropyReduction2021}.
	
	In the supercritical regime, the actual shape of the numerically obtained distributions depends on the distance from the threshold. First, the variances grow with $A/k_BT$, see also 
	Fig.~\ref{fig:moments_tau_015}(a). A second observation from Fig.~\ref{fig:distributions_heat_tau_015} is that, close to the threshold ($A/k_B T=20$), the distribution at the larger integration time is
	inclined to the right while far beyond the threshold ($A/k_BT=100$), it is rather inclined to the left. This is also seen from the third moments plotted in Fig.~\ref{fig:moments_tau_015}(b).
	At the larger integration time, the skewness crosses zero around $A/k_BT\approx 29.9$ (see dashed line). Interestingly, this value can be (approximately) related to the change of sign of a term in an expansion of the deterministic force around $\gamma v_\infty$, see Appendix~\ref{sec:corrections}. A similar argument can be made regarding the (even larger) values of $A/k_B T$, where the skewness starts to saturate to a constant value, see dashed-dotted line in Fig.~\ref{fig:moments_tau_015}(b).
	
	Finally, it is interesting to compare the distributions in the supercritical regime, Fig.~\ref{fig:distributions_heat_tau_015}(b)-(d),
	with the {\em Gaussian} (and thus, symmetric) distributions expected for a particle moving with the constant (deterministic) velocity $v_\infty$,
	corresponding to the constant (deterministic) force 
	$F=\gamma v_{\infty}$. The moments of the Gaussian are given in Eqs.~\eqref{eq:frist_moment_det} - (\ref{eq:moments_det}). We see that the Gaussians only give a rough description of the true distributions. Note that the (rather) small differences
	regarding the location of the maximum are expected already from corresponding differences in the heat production rate, $\langle\dot{q}\rangle$, see Fig.~\ref{fig:heat_rate_function_of_A}.
	As for the variance, we see from Fig.~\ref{fig:moments_tau_015}(a) that the Gaussian approximation becomes most reliable at large feedback strength and long integration times.
	%
	%
	%
	%%%%%%%%%%%%%%%%%%%%%%%%%%%%%%%%%%%%%%%%%%%%%%%%%%%%%%%%%%%%%%%%%%%%%%%%%%%%%%%%%%%%%%%%%%%%%%%%%%%%%%%%%%%%%%%%%%%%%%%%%%%%%%%%%%%%%%%%%%%%%%%
	%
	%
	%
	%
	%%%%%%%%%%%%%%%%%%%%%%%%%%%%%%%%%%%%%%%%%%%%%%%%%%%%%%%%%%%%%%%%%%%%%%%%%%%%%%%%%%%%%%%%%%%%%%%%%%%%%%%%%%%%%%%%%%%%%%%%%%%%%%%%%%%%%%%%%%%%%%%
	%\clearpage
\section{\label{sec:conclusions}Conclusions}

In this study we have investigated heat production in a
nonlinear stochastic system with time delay, using numerical and, in limiting cases, analytical techniques. 
From a physical point of view, main results of our study are the presence of nonzero average heat production $\langle\dot{q}\rangle$ already below the onset of persistent motion, and 
the complex parameter dependence of $\langle\dot{q}\rangle$ above the threshold. In particular, we find a pronounced maximum as function of the delay time, mirroring a corresponding maximum of the long-time velocity. 
In this aspect, our results are consistent with those of a recent experimental study of feedback-driven motion \cite{bell-davies_dynamics_2023}.

A large part of our paper has been devoted to the analysis of a linearized system as an approximation for small feedback strengths. We note that for the type of feedback force considered here, which contains a Pyragas-like displacement within one delay time, the linear stochastic delay equation has, strictly speaking, no stable stationary state. Rather we are just beyond the boundary of stability. Still, it turns out that for feedback parameters below a threshold (that coincides with that of persistent motion in the nonlinear case), the numerical calculations yield a finite value of $\langle\dot{q}\rangle$. 
This observation conforms with previous studies of linear non-Markovian systems at stability boundaries \cite{loos_heat_2019,loosMediumEntropyReduction2021,fernandezNonequilibriumDynamicsEntropy2024}. 
From a quantitative point of view, the linear approximation yields a useful upper bound of the heat production rate in the nonlinear case. Also, at small delay times, the numerical data coincide with that of a small delay approximation. 

Beyond the threshold, the interpretation (and benchmarking) of our numerical results was heavily based on the comparison with the corresponding deterministic problem. The latter has a clear-defined, even analytically accessible long-time solution \cite{kopp_persistent_2023-1}. It turns out that the deterministic dynamics indeed provides a fairly good description of the {\em average} heat production rate, at least at the noise levels considered.
Searching for such a solution could be a useful strategy also for other nonlinear feedback problems with time delay that possess a (nonequilibrium) steady state.

For future work, it might be interesting to investigate in more detail other thermodynamic quantities, particularly distribution functions and associated fluctuation relations \cite{mazonkaExactlySolvableModel1999}.
Again, exact relations for delayed systems are, so far, restricted to linear systems where important progress has recently been made \cite{rosinberg_stochastic_2017,rosinbergFluctuationsDynamicalObservables2024}.
Our results for the distribution of dissipated heat indicate that, even in the limit of weak feedback, the differences to the linear case are significant [see Fig.~\ref{fig:distributions_heat_tau_015}(a)].
Along with distributions, another interesting route of investigations would concern the so-called thermodynamic uncertainty relation \cite{baratoThermodynamicUncertaintyRelation2015,horowitzThermodynamicUncertaintyRelations2020}.
The question of which kind of systems do (or do not) fulfill this relation is still a challenging topic \cite{pietzonkaClassicalPendulumClocks2022}, especially in the non-Markovian case \cite{diterlizziThermodynamicUncertaintyRelation2020,platiThermodynamicBoundsDiffusion2023}.
Another interesting aspect, that we did not touch on, concerns the thermodynamic efficiency \cite{toyabeExperimentalDemonstrationInformationtoenergy2010,paneruLosslessBrownianInformation2018,admonExperimentalRealizationInformation2018} of feedback-driving with time delay, which is also interesting from an experimental point of view.

Finally, the results of the present paper could serve as a starting point to understand the nonequilibrium thermodynamics of the {\em collective} behavior in feedback-driven systems \cite{mijalkov_engineering_2016,piwowarczykInfluenceSensorialDelay2019,holubec_finite-size_2021,pakpourDelayinducedPhaseTransitions2024,tarama_traveling_2019,kopp_spontaneous_2023}. In particular, 
ensembles of feedback-driven particles of the type considered here show spontaneous large-scale velocity ordering, similar to Viscek-like systems \cite{kopp_spontaneous_2023}. The characterization of such transitions via thermodynamic notions is a research topic of current interest  \cite{borthneTimereversalSymmetryViolations2020a,grandpreEntropyProductionFluctuations2021,yuEnergyCostFlocking2022,ferrettiSignaturesIrreversibilityMicroscopic2022a}.

	%%%%%%%%%%%%%%%%%%%%%%%%%%%%%%%%%%%%%%%%%%%%%%%%%%%%%%%%%%%%%%%%%%%%%%%%%%%%%%%%%%%%%%%%%%%%%%%%%%%%%%%%%%%%%%%%%%%%%%%%%%%%%%%%%%%%%%%%%%%%%%%
	%
	%
	%
	%
	%%%%%%%%%%%%%%%%%%%%%%%%%%%%%%%%%%%%%%%%%%%%%%%%%%%%%%%%%%%%%%%%%%%%%%%%%%%%%%%%%%%%%%%%%%%%%%%%%%%%%%%%%%%%%%%%%%%%%%%%%%%%%%%%%%%%%%%%%%%%%%%
	\begin{acknowledgements}
		We thank Jan Meibohm for fruitful discussions and Alessio Zaccone for valuable comments on the potential interplay between time-dependent external forces and noise in the model. We gratefully acknowledge the support of the Deutsche Forschungsgemeinschaft (DFG, German Research Foundation), project number 163436311 - SFB 910.
	\end{acknowledgements}
	
	%%%%%%%%%%%%%%%%%%%%%%%%%%%%%%%%%%%%%%%%%%%%%%%%%%%%%%%%%%%%%%%%%%%%%%%%%%%%%%%%%%%%%%%%%%%%%%%%%%%%%%%%%%%%%%%%%%%%%%%%%%%%%%%%%%%%%%%%%%%%%%%
	%
	%
	%
	%
	%%%%%%%%%%%%%%%%%%%%%%%%%%%%%%%%%%%%%%%%%%%%%%%%%%%%%%%%%%%%%%%%%%%%%%%%%%%%%%%%%%%%%%%%%%%%%%%%%%%%%%%%%%%%%%%%%%%%%%%%%%%%%%%%%%%%%%%%%%%%%%%
	\appendix
	\section{Numerical methods}
	Simulations and data analysis were done using the Python programming language. The Python package matplotlib~\cite{Hunter:2007} was used for visualization.
	\subsection{\label{sec:num_sdde}Numerical solution of the Langevin equation}
\begin{figure}%[h!]
		\includegraphics[width=.3\textwidth]{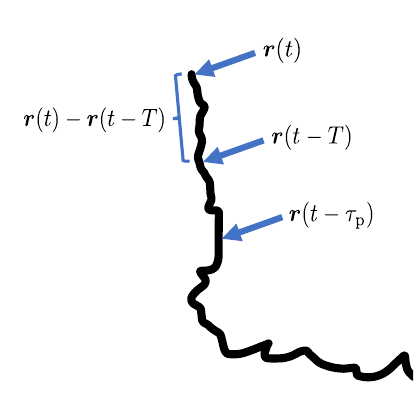}
		\caption{Illustration of the computation of the velocity based on trajectory data (see Eq.~\eqref{eq:v_app}) in the parameter regime of persistent motion.}\label{fig:velocity_computation}
	\end{figure}

	To solve the SDDE~\eqref{eq:sdde} numerically, we perform BD simulations based on the Euler-Maruyama integration scheme~\cite{maruyama_continuous_1955-2, kloeden_numerical_1992-4}.
	The discretized equation of motion reads
	\begin{align}
		\bm{r}_{n+1} &= \bm{r}_{n} + \gamma^{-1}\bm{F}(\bm{r}_{n}, \bm{r}_{n-N_\tau})\Delta t \\
		&~~+ \gamma^{-1}\sqrt{2k_B T \gamma \Delta t}~\bm{\eta}_{n+1},\nonumber
	\end{align}
	where $N_\tau$ is the number of time steps within one delay time $\tau$, $\Delta t$ is the size of the
	time step, and $\bm{\eta}$ is a vector of random numbers drawn from uncorrelated standard normal distributions.
	Note that an initial history has to be imposed for the time interval $[-\tau, 0]$ via the discretized history function $\bm{\Phi}(t_n) = \bm{\Phi}_{t_n}$, 
	i.e. the trajectory on the discretized time interval $[-\tau, 0]$ has to be known. 
	For all stochastic simulations in this study, we used trajectories of free Brownian particles as initial history. Indeed, we found that the specific choice for the initial history is not crucial for our numerical results, consistent with our observations in Ref.~[42]. 
	Further, we used time steps of $\Delta t = 10^{-4}\tau_B$ (in test calculations up to $\Delta t = 10^{-5}\tau_B$), to ensure sufficient numerical precision in all cases considered, where we specifically ensured that the  relation $\Delta t \ll \tau$ is upheld.
	Special care has to be taken when computing the magnitude of the velocity, $v$ (see Fig.~\ref{fig:vinfty_vs_A}).
	The reason is that, to obtain the magnitude, we have to square the velocity vector, involving noise correlations at zero time difference.
	However, it turns out that it is possible to calculate $v=|\bm{v}(t)|$ at feedback parameters where the particle performs persistent motion. 
	Specifically, we define the velocity according to
		\begin{equation}
			\bm{v}(t) = \frac{\bm{r}(t) - \bm{r}(t-T)}{T},
			\label{eq:v_app}
		\end{equation} 
	where the time $T$ must fulfill the condition $\Delta t \ll T < \tau_\mathrm{p}$, see Fig.~\ref{fig:velocity_computation} for an illustration.
	Here, $\Delta t$ is the simulation time step and $\tau_\mathrm{p}$ is the
	persistence time describing how long the particle moves, on average, along the same direction before noise leads to a reorientation \cite{kopp_persistent_2023-1}.
	In other words,  Eq.~(\ref{eq:v_app}) utilizes 
	the displacement within a time interval $T$ during which the particle moves (on average)  straight ahead. 
	Typically, $\tau_p$ is of the order of several $\tau$ while we keep $T$ fixed at $T=0.1\tau_B$.
	Having defined $\bm{v}(t)$ we evaluate $\sqrt{\bm{v}^2}$ and 
	perform an average over noise realizations (up to $N = 10^5$), yielding the magnitude of the velocity, $v=v(t)$. We have found that, in the long time limit, $v=\langle |\bm{v}| \rangle$ calculated in this way agrees well with
	the value obtained when fitting the mean-squared displacement of the feedback-driven particle to that of an active Brownian particle, as done in Ref.~\cite{kopp_persistent_2023-1}.
	For the numerical computation of the mean heat production rate $\langle\dot{q}\rangle$ (see Eq.~\eqref{eq:num_heat_rate}) we combine
	the average over noise realizations (up to $N=10^5$) with an average over $100$ simulation time steps to improve convergence and, thus, make numerical calculations feasible. We note that this does not affect the observed behavior (checked through test calculations).
	\subsection{Numerical treatment of Stratonovich calculus \label{sec:num_strat}}
	The discretized version of the heat production rate [Eq.~\eqref{eq:qdot}] in Stratonovich calculus \cite{kloeden_numerical_1992-4,gardinerStochasticMethodsHandbook2009} reads
	\begin{equation}
		\dot{q}(t_n) = \bm{F}(\bm{r}(t_{n}),\bm{r}(t_{n-N_\tau})) \cdot \left[\frac{\bm{r}(t_{n+1}) - \bm{r}(t_{n-1})}{2\Delta t}\right],
	\end{equation}
	where $\Delta t$ is the time step, $t_n = n\Delta t$ and $\bm{F}(\bm{r}(t_n),\bm{r}(t_{n-N_\tau})) = \gamma\dot{\bm{r}}(t_{n+1}) - \bm{\xi}(t_{n+1})$. 
	Directly related to the computation of $\dot{q}$ is the computation of the heat dissipated over a certain time interval, see Eq.~(\ref{eq:heatintegrated}).
	Following the Stratonovich convention, the discretized version of this integral (see, e.g.,~\cite{cates_stochastic_2022}) reads
	\begin{align}
		q(t_n) &= \sum_{n=1}^{N} \dot{q}(t_{n})\Delta t \\
		&= \sum_{n=1}^{N}  \bm{F}(\bm{r}(t_{n}),\bm{r}(t_{n-N_\tau})) \cdot \left[\frac{\bm{r}(t_{n+1}) - \bm{r}(t_{n-1})}{2\Delta t}\right]\Delta t.
	\end{align}
	\section{Derivation of Eq.~\eqref{eq:heat_rate_displacement}}\label{sec:linear_feedback_force}
\begin{figure}%[h!]
		\centering
		\includegraphics[width=.425\textwidth]{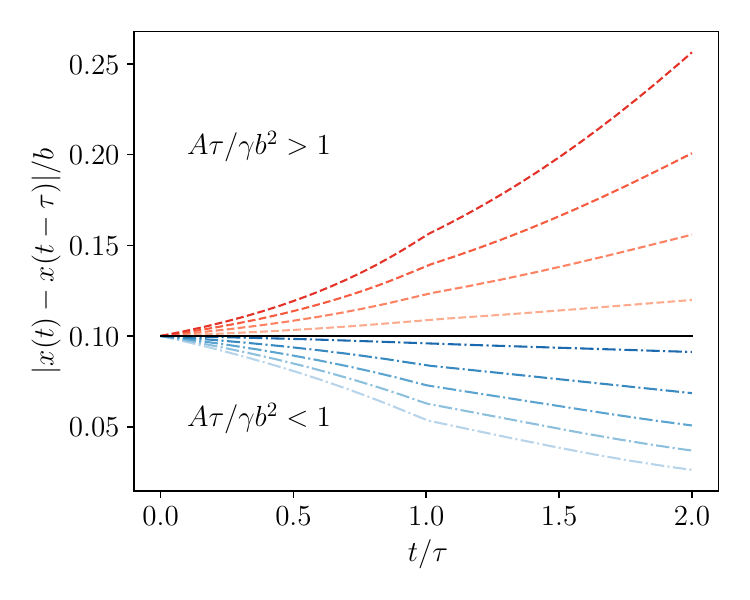}
		\caption{Displacement within one delay time according to the analytical solution \eqref{eq:analytical_solution_position1}-\eqref{eq:analytical_solution_position3}
				of the linear, one-dimensional DDE for different values of $A\tau/\gamma b^2$. Dashed red lines: results for $A\tau/\gamma b^2>1$, dashdotted blue lines: $A\tau/\gamma b^2<1$,		
				the solid line corresponds to the behavior at the threshold.
			The initial history function has the form $\Phi(t) = v_0 t$ on the interval $\left[-\tau,0\right)$ where $v_0$ was chosen to be constant based on $[x(0) - x(-\tau)]/b = 0.1$.
		} \label{fig:position_and_displacement}
	\end{figure}

	To derive Eq.~\eqref{eq:heat_rate_displacement} in the main text we start from Eq.~\eqref{eq:qdot} and consider the linear feedback force $\bm{F}_\mathrm{lin} = k_1 \bm{r}(t) - k_2 \bm{r}(t-\tau)$. 
	Plugging in the Langevin equation~\eqref{eq:sdde} 
	we obtain (suppressing the circle):
	\begin{align}
		\dot{q}_\mathrm{lin} 
		&= \frac{1}{\gamma} \left\lbrace \left[k_1 \bm{r}(t) - k_2\bm{r}(t-\tau)\right]^2\right.\nonumber\\
		&~~\left. + k_1 \bm{r}(t) \cdot \bm{\xi}(t) - k_2 \bm{r}(t-\tau) \cdot \bm{\xi}(t)\right\rbrace.
	\end{align} 
	Taking the noise average we find
	\begin{align}
		\langle\dot{q}_\mathrm{lin}\rangle &= \frac{1}{\gamma}\left\lbrace\langle\left[k_1 \bm{r}(t) - k_2\bm{r}(t-\tau)\right]^2\rangle\right. \nonumber\\
		&~~\left.+ k_1 \langle\bm{r}(t) \cdot \bm{\xi}(t)\rangle - k_2 \langle\bm{r}(t-\tau) \cdot \bm{\xi}(t)\rangle\right\rbrace.
	\end{align} 
	The equal-time correlation function $\langle\bm{r}(t) \cdot \bm{\xi}(t)\rangle$ can be calculated as described in \cite{frank_stationary_2001-2}. 
	For the correlation $\langle\bm{r}(t-\tau) \cdot \bm{\xi}(t)\rangle$, we use a causality argument~\cite{loos_heat_2019}, that is, 
	future noise cannot influence a past trajectory. We therefore have
	\begin{equation}
		\frac{1}{\gamma}\langle \bm{r}(t-\tau) \cdot \bm{\xi}(t) \rangle = 2D \delta_\tau,
	\end{equation}
	where we note that the right-hand side is valid in general and not only in a nonequilibrium-steady state~\cite{loos_heat_2019}.
	This finally yields Eq.~\eqref{eq:heat_rate_displacement}, with $k_1=k_2=A/b^2$.
	\begin{widetext}
	\section{Deterministic limit of displacements\label{sec:finite_displacements}}
		To understand why, in the linear case, displacements remain finite below the threshold $A \tau/\gamma b^2=1$, we consider the deterministic limit of the SDDE~\eqref{eq:sdde_linear}. 
		
		Linear delay-differential equations can be solved analytically, using the so-called `method of steps' (MOS)~\cite{driver_ordinary_1977,erneux_applied_2009-1}. If we prescribe a history function $\mathbf{\Phi}(t) = \bm{v}_0 t$ on the time interval $t\in[-\tau,0]$, where $\bm{v}_0$ points in $x$-direction, we obtain via the MOS the position as function of time (here up to $t=2\tau$)~\cite{kopp_persistent_2023-1}
		\begin{align}
			x(t) &= 	v_0 t, & -\tau < t \leq 0 \label{eq:analytical_solution_position1}\\
			x(t) &=	v_0\left[\frac{\gamma b^2}{A} + (t - \tau)  + \left(\tau - \frac{\gamma b^2}{A}\right)\exp\left(\frac{t A}{\gamma b^2}\right)\right], &  0 < t \leq \tau \label{eq:analytical_solution_position2} \\
			x(t) &=	v_0 \left\lbrace \frac{2 \gamma b^2}{A} + (t-2\tau) + \left( \tau-\frac{\gamma b^2}{A}\right)\exp\left(\frac{t A}{\gamma b^2}\right)\right. &\nonumber\\
			&\hphantom{=}\left.+ \left(t + \frac{\tau^2 A}{\gamma b^2} - \frac{t\tau A}{\gamma b^2} - \frac{\gamma b^2}{A}\right)\exp\left[\frac{\left(t-\tau\right) A}{\gamma b^2}\right]\right\rbrace, & \tau < t \leq 2\tau.\label{eq:analytical_solution_position3}
		\end{align}
	\end{widetext}
	Using the analytical expressions for the position (Eqs.~\eqref{eq:analytical_solution_position1} to~\eqref{eq:analytical_solution_position3}) we can now compute the displacement, $x(t) - x(t-\tau)$, within one delay time for $t \in (0,2\tau)$. The displacement as a function of time is plotted in Fig.~\ref{fig:position_and_displacement}.
	
	As depicted in Fig.~\ref{fig:position_and_displacement}, below the threshold $A\tau/\gamma b^2 = 1$ (dash-dotted blue lines), the displacement decreases as time goes on. 
	From solving the corresponding DDE numerically we know that this trend continues for long times and does not crucially depend on the initial history function we prescribe~\cite{kopp_persistent_2023-1}.
	The feedback force follows the same trend, due to its linear dependence on the displacement.
	This behavior resembles a trap that restricts the particle's motion and limits the occurring forces.
	Above the persistence threshold, displacements and thus feedback forces enter a positive feedback loop and quickly diverge, rendering the system unstable.
	\section{Corrections to the constant driving force approximation\label{sec:corrections}}
	The purpose of this Appendix is to investigate, within the "supercritical" regime of persistent motion, the dynamics of small perturbations of the true trajectory, $\bm{r}(t)$, from the deterministic one where the particle moves with 
	constant velocity, $\bm{v}_\infty$. We build on Ref.~\cite{kopp_persistent_2023-1} where we have used a similar analysis to explore the stability of the constant-velocity state in the deterministic limit.
	
	We here restrict ourselves, for simplicity, to a one-dimensional consideration (the relevant axis is that of the persistent motion in a short time interval). Also, noise effects are neglected.
	Following~\cite{kopp_persistent_2023-1}, we can derive an approximate equation of motion for the quantity $x(t)-v_\infty t$ by expanding the nonlinear feedback force in terms
	of $\Delta(t)=x(t) - x(t-\tau) - v_\infty \tau$. We assume that
	$\Delta(t)\ll b$.
	In \cite{kopp_persistent_2023-1} we have only considered the linear term (in the context of a linear stability analysis). We now take into account terms up to third order, yielding
	\begin{equation}
		\gamma\dot{x} \approx  \gamma v_\infty + K_1 \Delta(t) + K_2 \Delta^2(t) + K_3 \Delta^3(t).\label{eq:approx_const_vel}
	\end{equation}
	The right side can be interpreted as an approximation of the force.
	
\begin{figure}
		\centering
		\includegraphics[width=0.425\textwidth]{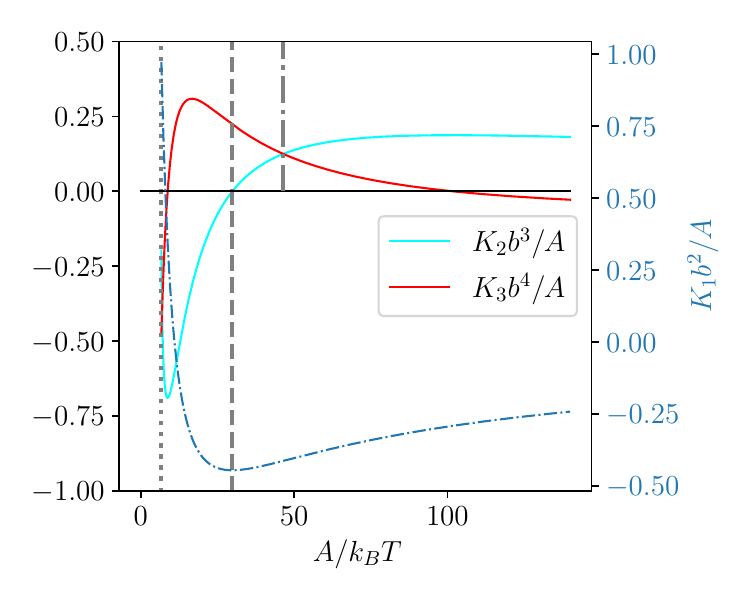}
		\caption{Left axis: coefficients $K_2$ and $K_3$ (solid lines) of the second and third order correction to the deterministic force, Eq.~\eqref{eq:approx_const_vel}.
				Right axis: coefficient $K_1$, corresponding to dashdotted blue line.
				Vertical dotted line: Persistence threshold ($A\tau/\gamma b^2 = 1$). Vertical dashed line: root of $K_2$.
				Vertical dash-dotted line: $K_2 = K_3$. Parameters: $\tau = 0.15 \tau_B$}.\label{fig:crossover_tau_015}
	\end{figure}

	The coefficient $K_1$ follows from the second derivative of the feedback potential in Eq.~\eqref{eq:nonlinear_potential} with respect to $x(t) - x(t-\tau)$ evaluated at $v_\infty \tau$, it reads~\cite{kopp_persistent_2023-1}
	\begin{equation}
		K_1= \frac{A}{ b^2}\exp\left[-\frac{\left(v_\infty \tau\right)^2}{2b^2}\right]\left[1-\frac{(v_\infty \tau)^2}{b^2}\right] .
	\end{equation}
	Similarly, we obtain
	\begin{align}
		K_2 &= \frac{v_\infty \tau}{2}\left[\frac{(v_\infty\tau)^2}{b^2} -3\right] \frac{A}{b^4} \exp\left[-\frac{\left(v_\infty \tau\right)^2}{2b^2}\right], \\
		K_3 &= \frac{1}{6}\left[-3 + \frac{6}{b^2}\left(v_\infty \tau\right)^2 - \frac{1}{b^4}\left(v_\infty\tau\right)^4 \right]\frac{A}{b^4} \exp\left[-\frac{\left(v_\infty \tau\right)^2}{2b^2}\right].
	\end{align}
	In Fig.~\ref{fig:crossover_tau_015} we plot the three coefficients $K_i$, $i=1,2,3$ as functions of $A/k_B T$ for a fixed delay time. It is seen that $K_1$ is essentially negative everywhere, while $K_2$ and $K_3$ can change their sign.
	The coefficient $K_2$ is particularly important when we assume that, in the stochastic case, the fluctuations of $x(t)-x(t-\tau)$ are symmetric around $v_\infty t$ and, thus, 
	$\langle\Delta\rangle$ and $\langle \Delta^3\rangle$ are negligible.
	
	For $6.66\leq A/k_B T\leq 29.87$, $K_2$ is negative. The corresponding term in Eq.~(\ref{eq:approx_const_vel}) thus yields a {\em negative} correction 
	to the constant force $\gamma v_\infty$. This may explain the fact that the average heat production rate, which involves the square of the force, is consistently {\em smaller} than in the constant-force limit in this regime, see
	Fig.~\ref{fig:heat_rate_function_of_A}. Upon increase of $A/k_BT$, $K_2$ then crosses zero at $A/k_B T = \gamma b^2 \exp\left(3/2\right)/\tau k_B T\approx 29.88$.
	Neglecting contributions proportional to $\langle\Delta\rangle$ and $\langle \Delta^3\rangle$, $K_2=0$ implies that the approximate force following from Eq.~(\ref{eq:approx_const_vel}) reduces to the constant term $\gamma v_\infty$.
	For such a system, the distribution of dissipated heat is Gaussian and thus, symmetric. This may explain the vanishing of the skewness at this feedback strength seen in Fig.~\ref{fig:moments_tau_015}(b).
	Finally, $K_2$ becomes equal to $K_3$ at $A/k_B T\approx 46.46$. For even larger feedback strengths, $K_2$ saturates to a constant positive value.
	This implies, overall, an increase of the force relative to $\gamma v_\infty$. Indeed, the mean heat production rate becomes larger than the deterministic limit, see Fig.~\ref{fig:heat_rate_function_of_A}.
	%
	%\section{References}
	\bibliography{Stochastische_Thermodynamik}% Produces the bibliography via BibTeX.
\end{document}